\documentclass[twocolumn,prd,groupedaddress,nofootinbib]{revtex4}

\begin{document}

\title{Quantum Fields in anti-de Sitter space and the Maldacena conjecture}
\author{Nelson R. F. Braga}
\email{braga@if.ufrj.br}
\affiliation{Instituto de F\'{\i}sica\\
Universidade Federal do Rio de Janeiro, RJ 21945-970 -- Brazil}

%\date{\today}

\begin{abstract}
We  review  in this lecture the relation between the Maldacena Conjecture,
also  known as $AdS/CFT$ correspondence, and  the so called 
Holographic principle that seems to be an essential ingredient for a 
quantum gravity theory. We  also illustrate the idea of Holography by showing 
that the curvature of the anti-de Sitter space reduces the number of degrees
of freedom making it possible to find a mapping between a quantum theory 
defined on the bulk and another defined on the corresponding boundary.
\end{abstract}

\pacs{98.80.Bp, 98.80.Cq}

\maketitle

\section{Introduction}
The interest of theoretical physicists in studying quantum fields in
anti de Sitter (AdS) space is not new\cite{Fro}.
In particular the question of the quantization of fields in this space
circunventing the problem of the lack of a Cauchy surface
was addressed in\cite{QAdS1,QAdS2}. 
There was, however, a remarkable increase in the attention devoted
to this subject since the appearance of  the Maldacena 
conjecture\cite{Malda} (see \cite{Malda2,Pe} for reviews)  on the equivalence 
(or duality) of the large $N$ limit of $SU(N)$ superconformal 
field theories in $n$ dimensions and supergravity and 
string theory in anti de Sitter spacetime in $n+1$  dimensions
also called AdS/CFT correspondence.
This correspondence was elaborated by  Gubser, Klebanov and Polyakov \cite{GKP}
and by Witten \cite{Wi} interpreting the boundary values of bulk fields as
sources of boundary theory correlators.

The large scientific impact caused by Maldacena work can be associated with
(at least) two reasons: 

\begin{itemize}
\item{(i)} It represents an explicit example of a holographic mapping between 
two quantum theories that live in different dimensions. That means: a realization of 
the  {\bf Holographic Principle}.

\item{(ii)} It also represents an important step  in the search for 
a string theory description of  QCD that would lead to a formulation for QCD 
at low energies (where perturbative calculations can not be used because of the 
strong coupling)
\end{itemize}

\bigskip 
As a remark let us mention that the large N limit of SU(N) gauge theories was studied
by 't Hooft a long time ago. He considered an expansion in the parameter $ 1/N$ 
were QCD gets a simpler form (planar digrams dominate $\,\rightarrow\,$ topological 
structure like strings) \cite{to}.
 
\bigskip
\section{Holographic principle}
The original motivation for the holographic principle was the study 
of black hole entropy. Let us start with the following question:  What happens with 
the entropy of the Universe when some portion of matter (say a rocket or anything else)
is absorbed by a black hole?
Classicaly (we mean without quantum mechanics) black holes can only absorb particles.
So, one could not associate any temperature with them. Thus it would make no sense to 
think that the black hole itself has some entropy and the entropy of the Universe 
would not be conserved when we throw some matter inside it.

Quantum effects however change this picture. The thermal radiation found out by 
Hawking lead himself and Bekenstein\cite{Be,Ha} to develop a  thermodynamical model
for  black holes. Based on the analysis of the dynamics of black holes they found 
a generalised form of the second law of thermodynamics where a black hole has 
an entropy proportional to area of its horizon

\bigskip

\begin{equation}
\label{bhe}
S_{BH}\,=\,{A \over 4\hbar G} 
\end{equation}

\bigskip

\noindent The generalized form of the second law of thermodynamics reads:

\bigskip

$\,\,\,\,\,\,\,\,\Delta S_{BH} \,+\, \Delta S_{Rest \,of\, Univ.}  \ge \,0 \,\,$

\bigskip
One should be able to find a way of calculating this entropy by summing up degrees 
of freedom of black holes. This is still an open problem because it would
depend essentially on the formulation of a satisfactory quantum theory for gravity. 
There are however some important results in 
this direction for example calculating black hole entropy using string theory as 
refs. \cite{BHE1,BHE2}.

An interesting consequence of eq. (\ref{bhe}) is that it contrasts with our standard
idea (without taking gravity into account) that the entropy 
is an extensive quantity. That means: entropy should be proportional to the  
volume not to the area.  This result lead 't Hooft and Susskind to formulate 
the Holographic  Principle:

{\it `` Physics of a quantum system with gravity in some \break volume V
can be described in terms of degrees of \break freedom that can be contained
in its boundary".}

\bigskip

\noindent  The general idea is that some systems tend naturally to black holes and
for the others we can find processes that transform them into black holes and increase 
their entropy. So: the maximum entropy is limited by the boundary area (in Planck units). 
That means:  Quantum mechanics plus gravity in 3 spatial dimensions is equivalent 
to an image that can be mapped  in a bidimensional projection.
The area of the boundary in units of  Planck area represents 
the  maximum number of degrees of freedom in the interior volume.
It is presently believed that a (candidate to ) quantum gravity theory
should satisfy this principle.

\vskip 1cm

\section{Implementation of Maldacena conjecture}

\bigskip
Soon after Maldacena's seminal article, Gubser , Klebanov and Polyakov\cite{GKP} and
Witten\cite{Wi} have elaborated the general conjecture showing  how to calculate
Physical quantities of a conformal theory  in the boundary of an anti-de Sitter ($AdS$)
space in terms of a bulk theory. In order to see how this correspondence holds,
let us first remind that an anti-de Sitter space of 
$n+1$ dimensions $\,(\,AdS_{n+1}\,)\,$ is a space of constant negative curvature  
that  can be taken as a Hyperboloid in a larger $n+2$ dimensional flat space
with coordinates  $(X^0,X^1,...,X^n,X^{n+1})$ and metric 
$\eta_{ab}={\rm diag}(+,-,-,...,-,+)$: 

\vskip 1cm

$$
(X_0)_2+(X_{n+1})^2-\sum_{i=1}^n(X_i)^2 = \Lambda^2={\rm constant}\,,
$$
 
\bigskip

\noindent  We can introduce coordinate systems inside $AdS$ like the global coordinates
 $\,\rho,\tau,\Omega_i\,$, more used before the discovery of the $AdS/CFT$ correspondence,
defined by:

\begin{eqnarray}
X_0 &=& \Lambda \,\sec\rho\, \cos \tau \nonumber\\
X_i &=& \Lambda \,\tan \rho\, \,\Omega_i\;\qquad\qquad
 \nonumber\\
X_{n+1} &=& \Lambda \sec \rho \,\sin\tau \nonumber
\end{eqnarray}

\noindent where $\sum_{i=1}^n\Omega^2_i=1\;,
0\le \rho <\pi/2\,\, ,\,\,0\le\tau< 2\pi$
 
\bigskip
\noindent Note that the time variable $\tau $ is compact, so we 
must ``unwrap" it by actually considering 
the  $AdS$ covering space (that means an infinite set of copies of $AdS$ spaces in the
$\tau$ direction.

For the   AdS/CFT correspondence the so called  Poincar\'e coordinates 
$(z\,,\,x^i\,,\,t\,)\,\,\,\,\,
$ ( with \,\,$z \ge 0 )$ are more usefull.
They are defined by the relations  

\bigskip

\begin{eqnarray}
X_0 &=& \quad {1\over 2z}\,\Big( \,z^2\,+\,\Lambda^2\,
+\,{\vec x}^{\,2}\,-\,t^2\,\Big)
\nonumber\\
X_i &=& \quad {\Lambda x^i \over z}
\nonumber\\
X_n &=& - \; {1\over 2z}\,
\Big( \,z^2\,-\,\Lambda^2\,+\,{\vec x}^{\,2}\,-\,t^2\,\Big)
\nonumber\\
X_{n+1} &=& \quad\; {\Lambda t \over z}
\nonumber
\end{eqnarray}

\noindent and the corresponding measure reads

$$
ds^2= { \Lambda^2 \over z^2 } \Big(\, (dz)^2 + (d\vec x)^2\,-\, (dt)^2 \,\Big)
$$

\bigskip

\noindent The $AdS$ boundary, where the Conformal Field Theory is defined
corresponds to the region $ z \,=\,0\,$ plus a point at infinity:  $ z   = \infty $.  

\bigskip

\noindent Witten\cite{Wi} has interpreted the $AdS/CFT$ correspondence in terms of a 
holographyc mapping:  {\it boundary values of fields that have dynamics defined  
inside the AdS space act as sources of correlation functions   
of the boundary conformal field theory (CFT)}. 
The simplest illustrative example is that of a scalar massless field (massless) 
in $AdS$ (more details can be found in \cite{GKP,Wi,FMMR,MV}. Taking a scalar field
in the bulk, with boundary value

\bigskip
 $\phi (z\,,\,x^i\,,t)\,\,\rightarrow\,\,
\phi_0 (\,x^i\,,t )\,\,\,\,$  
as  $\,\,z\,\rightarrow\,0\,$.

\vskip1cm

\noindent and conformal operators ${\cal O}\,(\,x^i\,,t )$ on the boundary, 
the generator of correlation functions for this  CFT operators
(now: $x^i,t \equiv x$ for simplicity) 
$$
Z\,[\,\phi_0 \,]\,=\, \langle \,\,\exp \int d^{n}x\; \phi_0 (\,x\,)\,
{\cal O} (\,x\,)\,\,\rangle\,\,
\nonumber
$$

\noindent such that

$$ 
{\delta\over \delta \phi_0 (\,x_1\,) } {\delta\over \delta \phi_0 (\,x_2\,) }
Z\,[\,\phi_0\,]\,=\,\langle \, { \cal O } (\,x_1\,){\cal O} (\,x_2\,)\,\rangle\,
$$
 
\bigskip

\noindent  where the fields $\phi_0 $ act as {\cal sources} will be calculated from
the on shell  action $I (\phi )$ for the bulk scalar field 
 $\,Z [\,\phi_0\,]\,\,\Longleftrightarrow\,\, exp \{ - I (\phi )\}  $

Considering the action for a scalar field in curved spacetime

$$L\,=\,{1\over 2} \int d^{n+1}x \sqrt{g} \,\partial_\mu \phi \,
\partial^\mu \phi
\,\,,\nonumber
$$

\noindent with equation of motion 

$$
\label{motion}
\nabla_\mu \nabla^\mu \phi\,=\,{1\over \sqrt{g}} \partial_\mu 
\Big( \, \sqrt{g} \partial^\mu \phi \,\Big) 
\,=\,0\,\,.
$$

\vskip1cm
\noindent we can relate the field in AdS bulk in terms of boundary values:

$$
\label{solution}
\phi ( z\,,\,x \,)\,=\,c\,\int d^n x^{\prime} 
{ (z)^n \over \left( (z)^2 \,+\,
( x \,-\,{ x}^\prime \,)^2\,\right)^n} 
\phi_ 0 ({x}^\prime )\,\,.
$$

\bigskip
\noindent and write the on shell action as

$$
I [\phi ] \,=\, - {c n\over 2} \,\int d^n x \,d^n x^\prime 
{\phi_ 0 ({x}^\prime )
\phi_ 0 ({x} ) \over ( x \,-\,{x}^\prime \,)^{2n}}
$$

\bigskip

\noindent So we find, for example, the two point correlation function for the operators  
${\cal O} ( x )\,\,$ 

\bigskip

$$ \langle {\cal O} ( x ) {\cal O} ( y \,) \rangle \,\,\sim
\,\,{ 1 \over ( x \,-\,y \,)^{2n}  }\,\,\,$$ 

\bigskip

\noindent as expected from conformal invariance.
 
\vskip 1cm
\noindent Let us remark that the operators  ${\cal O}$ of the CFT are not  
in general scalar fields but rather composite operators whose conformal  dimension
depends on the  dimension of the space.

\section{Counting degrees of freedom in $AdS$ space}

\noindent Considering the geometry of $AdS$ space in Poincare \break coordinates 
we see that a hypersurface of area $\,\Delta A \,$ 
corresponding to $\,\Delta x^1 \, ... \, \Delta x^{n-1} \,$
at some small $z = \delta $ will generate a finite volume $\Delta V$ if we 
move it along $z$ till 
$z \rightarrow \infty $

\begin{eqnarray}
\Delta A &=& \int \frac {d^{n-1}x}{(z/\Lambda)^{n-1}}
= \left(\frac{\Lambda}{\delta}\right)^{n-1}
\,\Delta x^1 \, ... \, \Delta x^{n-1}
\nonumber\\
\nonumber\\
&\Delta V & \,=\,\int \frac {d^{n-1}x\,dz}{(z/\Lambda)^{n}}
= \Lambda \,{\Delta A \over n-1}.
\nonumber
\end{eqnarray}

\noindent This fact is illustrated in the figure bellow.
\vskip 1cm

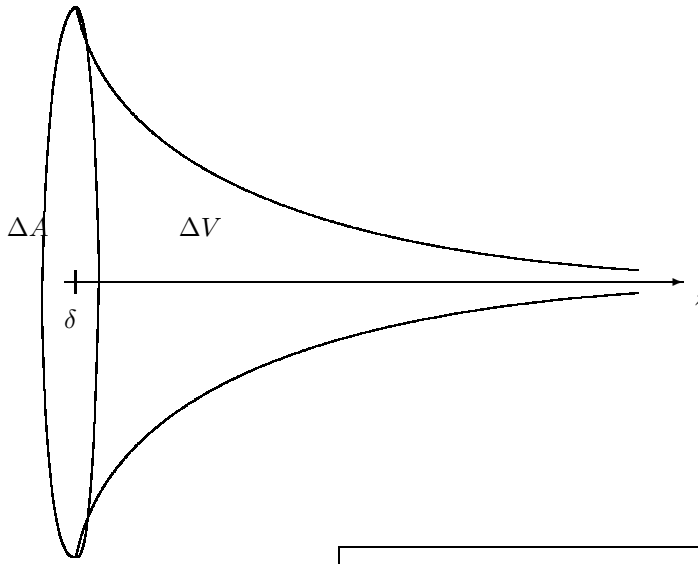
\begin{widetext}

\vskip 2cm
\vskip 1cm

\hskip 2cm
%%%%%%%%%%%%%%%%%%%%%%%%%%%%%%%%%%%%%%%%%%%%%%%%%%%%%%%%%%%%
%%%%%%%%%%%%%%%%%%%%%%%%  Figure 1%%%%%%%%%%%%%%%%%%%%%%%%%%
%%%%%%%%%%%%%%%%%%%%%%%%%%%%%%%%%%%%%%%%%%%%%%%%%%%%%%%%%%%%
\
\setlength{\unitlength}{0.06in}
\begin{picture}(50,10)(-20,0)
\label{warpfig}
\rm
%%%%%%%%%%%%%%%%%%%  Surface %%%%%%%%%%%%%%%%%%%%%%%%%%%%%%%%

\bezier{1000}(1,25)(-1.5,24)(-2,-1)
\bezier{1000}(-2,-1)(-1.5,-24)(1,-23)
\bezier{1000}(1,-23)(2.5,-24)(3,2)
\bezier{1000}(3,2)(2.5,24)(1,25)
\put(-5,5){$\Delta A$}
\put(10,5){$\Delta V$}
%%%%%%%%%%%%%%%%%%%Eixo Horizontal%%%%%%%%%%%%%%%%%%%%%%%%%%%%
\put(0,1){\vector(1,0){54}}
\put(1,0){\line(0,1){2}}
\put(0,-3){$\delta$}
\put(55,-1){$z$}
%%%%%%%%%%%%%%%%%%%%Dados%%%%%%%%%%%%%%%%%%%%%%%%%%%%%%%%%%%%%%
\bezier{800}(1,25)(5,5)(50,2)
\bezier{800}(1,-23)(5,-3)(50,0)
\end{picture}
\vskip 1cm
%%%%%%%%%%%%%%%%%%%%%%% end of figure 1 %%%%%%%%%%%%%%%%%%%%%%%%
%%%%%%%%%%%%%%%%%%%%%%%%%%%%%%%%%%%%%%%%%%%%%%%%%%%%%%%%%%%%%%%%

\vskip 2cm

\bigskip
\end{widetext}

\noindent If we associate (in a regularised way) degrees of freedom  with volume cells 
of the space we see that we can map the $\Delta V$ cells in  area cells
of $\Delta A$.
This reduction of degrees of freedom 
associated with the fact that the volume is proportional to the
area should be reflected in a quantum theory in $AdS$ $\,$.
How can it be so?
The hint comes from the analisys of the Cauchy problem in the $AdS$ space
considered in \cite{QAdS1,QAdS2}: 
massless particles can enter or left the $AdS$ from spatial infinity in finite times.
So a consistent quantization (that means a well defined Cauchy problem) requires 
a compactification of the space. 
One has to add a surface at infinity where boundary conditions have to be imposed.
(Like the cover of a box: nothing enters or lefts the space).
These references considered global coordinates where 
the boundary is the hypersurface $\,\rho = \pi/2$.
 
Now considering the Cauchy problem in Poincar\'e Coordinates, particles without mass 
may come from or go to $\,z\, \rightarrow\,\,\infty \,$ in a finite time.
But how can we compactify the z coordinate that has infinite range?
The answer \cite{BB1,BB2,BB3} is that one needs to introduce another coordinate chart
$(\,z^{\,\prime} , x^i , t ) \,$ related to the original one by

\begin{eqnarray}
{1\over z^\prime} = {1\over \delta} - {1\over z} 
\end{eqnarray}

\noindent were we take the range of the coordinates $z$ as $\delta \le z \le R \,$ 
in order to avoid  the singularity at the origin and to stop the first chart at 
some arbitrarily large $R$. For $z^\prime $ we take  and 
$\delta \le z^\prime \le R^\prime \,$   where $R^\prime= \delta R /( R - \delta )\,$ 
in order to map the rest of the space. The two sets must be compact because we need to 
impose boundary conditions that match them togheter.

Now the "point at infinity" in the $z$ coordinate is mapped at $z^\prime = \delta$

\bigskip
 \noindent 
The important consequence of this representation of the compact $AdS$ space in 
terms of two compact coordinate charts is that even in the Poincare coordinate chart
the field spectrum is {\bf discrete}

\begin{widetext}

\begin{equation}
\label{QF}
\Phi(z,\vec x,t)\,=\, 
 = \sum_{p=1}^\infty \,\, \int {d^{n-1}k \over (2\pi)^{n-1}}\,
{z^{n/2} \,J_\nu (u_p z ) \over R w_p(\vec k ) 
\,J_{\nu\,+\,1} (u_p R ) }
\lbrace { a_p(\vec k )\ }
 e^{-iw_p(\vec k ) t +i\vec k \cdot \vec x}\,
\,+\,\,c.c.\rbrace
\end{equation}

\end{widetext}
 
\bigskip

\noindent The important consequence of this discretization of the field spectrum is 
that it makes possible to
find a one to one mapping between bulk and boundary theories.
This would not be possible if the spectrum were continous.
We can understand this by an analogy with the fact that it is possible to find 
a one to one mapping of an enumerable set of lines 
into one single line but it is not possible to map (one to one) a plane and a line.
This is explained in \cite{BB4} were an explicit mapping between scalar fields 
in $AdS$ space and scalar fields on the boundary is presented.

\end{document}